\documentclass[twoside,11pt]{article}

\usepackage{hyperref}
\usepackage{amsmath}
\usepackage{jmlr2e}

\jmlrheading{}{}{}{}{}{}

\ShortHeadings{Homogeneous temporal activity patterns in a large online communication space}{}
\firstpageno{1}

\begin{document}

\title{Homogeneous temporal activity patterns in a large online communication space}

\author{\name Andreas~Kaltenbrunner \email andreas.kaltenbrunner@upf.edu \\
        \name Vicen\c{c}~G\'omez \\
        \name Ayman Moghnieh  \\
        \name Rodrigo Meza  \\
        \name Josep Blat \\
        \name Vicente~L\'opez \\\\
        \addr Departament de les Tecnologies de la Informaci\'o i les
comunicacions\\
Universitat Pompeu Fabra\\
Passeig de Circumval{\textperiodcentered}laci{\'o} 8, 08003 Barcelona, Spain\\\\
Barcelona Media Centre d'Innovaci{\'o}\\
Ocata 1, 08003 Barcelona, Spain}
\maketitle

\begin{abstract}
The many-to-many social communication activity on the popular
technology-news website Slashdot has been studied. We have
concentrated on the dynamics of message production without considering
semantic relations and have found regular temporal patterns in the
reaction time of the community to a news-post as well as in single
user behavior. The statistics of these activities follow log-normal
distributions.  Daily and weekly oscillatory cycles, which cause
slight variations of this simple behavior, are identified.  A
superposition of two log-normal distributions can account for these
variations.  The findings are remarkable since the distribution of the
number of comments per users, which is also analyzed, indicates a
great amount of heterogeneity in the community.  The reader may find
surprising that only a few parameters
allow a detailed description, or even prediction, of social
many-to-many information exchange in this kind of popular public
spaces.
\end{abstract}


\section{Introduction}

\label{sec:intro}\noindent
Nowadays, an important part of human activity leaves electronic traces
in form of server logs, e-mails, loan registers, credit card
transactions, blogs, etc.  This huge amount of generated data allows
to observe human behavior and communication patterns at nearly no cost
on a scale and dimension which would have been impossible some decades
ago.  A considerable number of studies have emerged in recent years
using some part of these data to investigate the time patterns of
human activity.  The studied temporal events are rather diverse and
reach from directory listings and file transfers (FTP requests)
\citep{Paxson95}, job submissions on a supercomputer~\citep{Kleban03},
arrival times of consecutive printing-job submissions \citep{Harder06}
over trades in bond \citep{Mainardi00} or currency futures \citep{Masoliver03}
to messages in Internet chat systems \citep{Dewes03}, online
games~\citep{Henderson01}, page downloads on a
news site~\citep{Dezso06} and e-mails~\citep{johansen04}.  A common
characteristic of these studies is that the observed probability
distributions for the waiting or inter-event times are heavy tailed.
In other words, if the response time ever exceeds a large value, then
it is likely to exceed any larger value as well~\citep{Sigman99}. A
recent study~\citep{barabasi05} tries to explain this behavior under
the assumption that these heavy tailed distributions can be well
approximated by a power-law or at least by a power-law with an
exponential cut-off~\citep{Newman05}. The cited study presents a model
which seems to explain the distribution of e-mail response times and
has been used later to account for the inter-event times of
web-browsing, library loans, trade transactions and correspondence
patterns of letters~\citep{Vazquez06}.  However, the hypothesis of a
power-law distribution is not generally accepted, at least in case of
e-mail response times.  \citet{Stouffer06} claim
that the data can be much better fitted with either a log-normal (LN)
distribution~\citep{LimpertSA01} or the superposition of two LN. This
debate has been repeated across many areas of science for decades, as
noticed by~\citet{mitzen2004}.

To the authors' knowledge no study of this type has been performed on
systems where social interaction occurs in a more complex manner than
just person to person (one-to-one) communication.  We think it is
valuable to analyze the temporal patterns of the many-to-many social
interaction on a technology-related news-website which supports user
participation. We have chosen
Slashdot\footnote{http://www.slashdot.org}, a popular website for
people interested in reading and discussing about technology and its
ramifications.  It gave name to the ``Slashdot
effect''~\citep{Adler1999}, a huge influx of traffic to a hosted link
during a short period of time, causing it to slow down or even to
temporarily collapse.

Slashdot was created at the end of 1997 and has ever since
metamorphosed into a website that hosts a large interactive community
capable of influencing public perceptions and awareness on the topics
addressed.  Its role can be metaphorically compared to that of
commercial malls in developed markets, or hubs in intricate large
networks.  The site's interaction consists of short-story
\textbf{posts} that often carry fresh news and links to sources of
information with more details. These posts incite many readers to
\textbf{comment} on them and provoke discussions that may trail for
hours or even days. Most of the commentators register and comment
under their nicknames, although a considerable amount participates
anonymously.

Although Slashdot allows users to express their opinion freely,
moderation and me\-ta-mod\-er\-a\-tion mechanisms are employed to
judge comments and enable readers to filter them by quality.  The
moderation system was analyzed by~\citet{Lampe04}
who concluded that it upholds the quality of discussions by
discouraging spam and offensive comments, marking a difference between
Slashdot and regular discussion forums.  This high quality social
interaction has prompted several socio-analytical studies about
Slashdot. \citet{Poor05} and \citet{Baoill00} have
both conducted independent inquiries on the extent to which the site
represents an online public sphere as defined by~\citet{Habermas89}.

Given that a great amount of users with different interests and
motivations participates in discussions about very different topics,
one would expect to observe a high degree of heterogeneity on a site
like Slashdot.  However, what if the posts and comments were analyzed
just as imprints of an occurring information exchange, with no regard
to semantic aspects?  Is there a homogeneous behavior pattern
underlying heterogeneity?  To answer these and related questions we
collected and studied one year's worth of interchanged messages along
with the associated meta-data from Slashdot.  We show here that the
temporal patterns of the comments provoked by a post are very similar,
indicating that homogeneity is the rule not the exception.  The
temporal patterns of the social activity fit accurately log-normal
distributions, thus giving empirical evidence of our hypothesis and
establishing a link with previous studies where social interaction
occurs in a simpler way.

Finally, our analysis allows more insight into questions such as: is
there a time-scale common to all discussions, or are they scale-free?
What does incite a user to write a comment, is it the relevance of the
topic, or maybe just the hour of the day?  Can we predict the amount
of activity a post will trigger already some minutes after it has been
written?  Which type of applications can we devise on the basis of
using these conclusions?

The rest of the article is organized as follows: In
section~\ref{sec:methods} we briefly explain the process of data
acquisition.  We then present the results in section~\ref{sec:results}
providing first an overview of the global activity and then explaining
our analysis in detail.  We finish the paper with
section~\ref{sec:discussion} where we discuss the results.

\section{Methods}
\label{sec:methods}\noindent
In this section we explain the methods used to crawl and analyze
Slashdot.  The crawled\footnote{Software used: wget, Perl scripts, and
  Tidy on a GNU/Linux, Ubuntu 6.0.6 OS.}  data correspond to posts and
comments published between August $26$th, $2005$ and August $31$th,
$2006$.  We divided the crawling process into two stages.  The first
stage included crawling the main HTML (posts) and first level comments
and the second stage covered all additional comment pages.  Crawling
all the data took $4.5$ days and produced approximately $4.54$ GB of
data.  Post-processing caused by the presence of duplicated comments
was necessary (due to an error of representation on the website).
Although a high amount of information was extracted from the raw HTML
(sub-domains, title, topics, hierarchical relations between comments)
we concentrated only on a minimal amount of information: \textbf{type}
of contribution (either post or comment), its \textbf{identifier},
\textbf{author}'s identifier and \textbf{time-stamp} or date of
publishing.  The selected information was extracted to XML-files and
imported into Matlab where the statistical analysis was performed.
Table \ref{table:main} shows the main quantities of the crawling and
the extracted data.  \vspace{-12pt}
\begin{table}[!hb]
\centering
\caption{Main quantities of crawling and retrieved data. \vspace{6pt}}
\begin{tabular}{lr}
  Period covered            & $26$-$8$-$05$ $-$ $31$-$8$-$06$\\
  Time needed for crawling   & $4.5$ days\\
  Amount of data mined       & $4.54$ GB\\ 
  Posts                 & $10016$ \\ 
  Comments              & $2075085$ \\ 
  Commentators          & $93636$ \\ 
  Anonymous comments         & $18.6\%$
\end{tabular}
\label{table:main}
\end{table}

The time-stamps of post and comments can be obtained from Slashdot
with minute-precision and corresponded to the EDT time zone ($=$
GMT$-4$ hours).  They allow to calculate the following two quantities:

The \textbf{Post-Comment-Interval (PCI)} stands for the difference
between the time-stamps of a comment and its corresponding post.

The \textbf{Inter-Comment-Interval (ICI)} refers to the difference
between the time-stamps of two consecutive comments of the same user
(no matter what post he/she comments on).

\section{Results}
\label{sec:results}
\noindent
In this section we first give an overview of the global activity
looking at the data on different temporal scales and analyzing some
relations between variables of interest.  We then focus on the
activity provoked by single posts and analyze the behavior of single
users, concentrating on the most active ones.

\subsection{Global cyclic activity}
\label{sec:global}
\noindent
As previously explained, comments can be considered as reactions
triggered by the publishing of posts. This difference in nature
between both types of contributions justifies a separate analysis of
their dynamics.

\begin{figure}[!tb]\centering
\includegraphics[angle=-90,width=0.48\columnwidth]{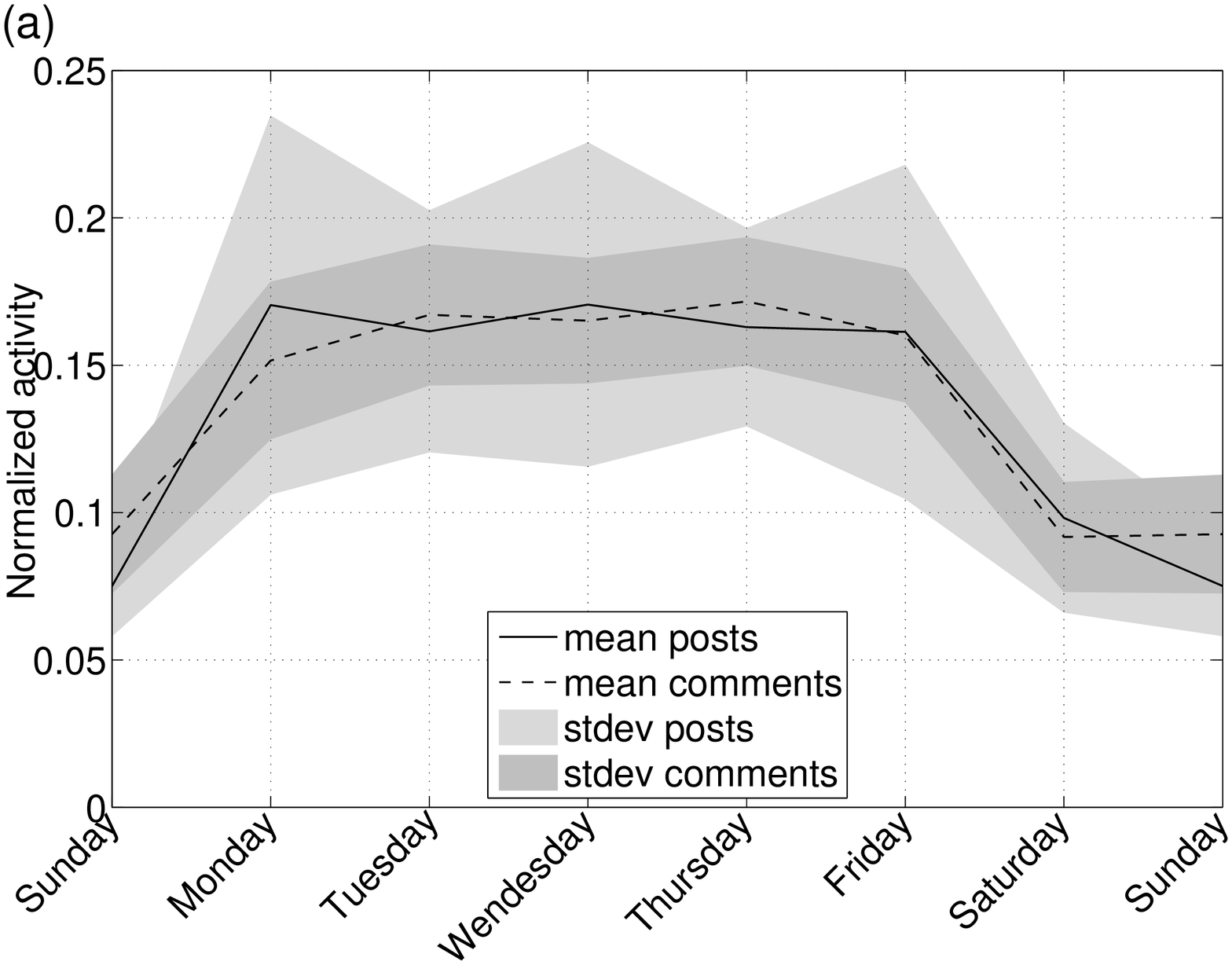}
\includegraphics[angle=-90,width=0.48\columnwidth]{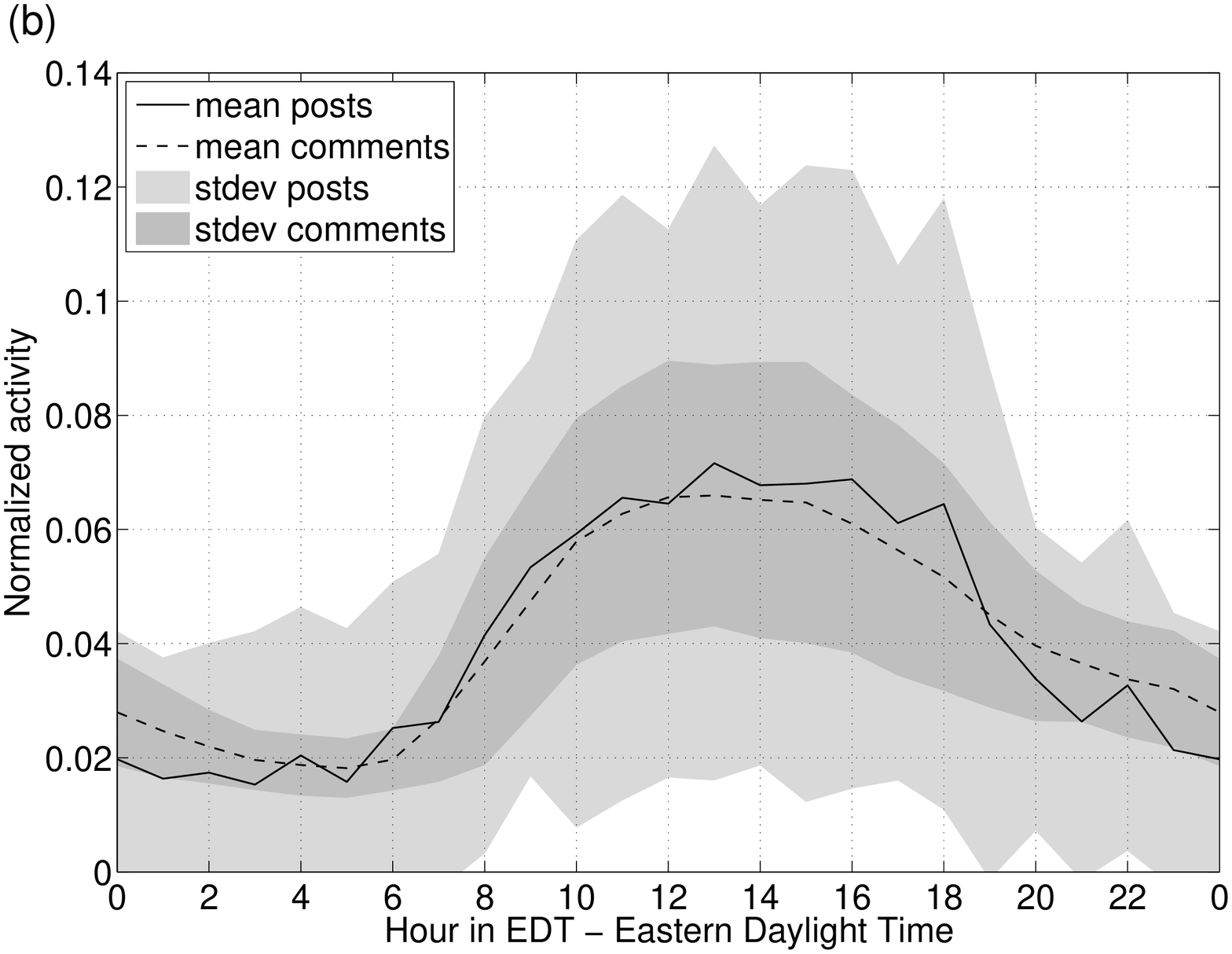}
\caption{\textbf{(a)} Weekly and \textbf{(b)} daily activity cycles.}
\label{fig:activity-week}
\label{fig:activity-hour}
\end{figure}

Figure \ref{fig:activity-week} shows (normalized) mean activity and
standard deviations of both posts and comments.  It illustrates
patterns in agreement with the social activity outside the public
sphere.  Figure~\ref{fig:activity-week}a shows regular, steady
activity during working days which slows down during weekends.  This
weekly cycle is interleaved by daily oscillations illustrated in
Figure~\ref{fig:activity-hour}b.  The daily activity cycle reaches its
maximum at $1$pm approximately and its minimum during the night
between $3$am and $4$am.  Although Slashdot is open to public access
around the world, we see that its activity profile is clearly biased
towards the American time-schedule.

Interestingly, although post activity shows more fluctuations and
higher standard deviations than comment activity, there is little
discrepancy between their mean temporal profiles.  This difference in
the deviations is not surprising given the greater number of comments
(see Table~\ref{table:main}).  We notice that the standard deviations
of the daily post- and commenting activities also show similar cyclic
behavior (Figure~\ref{fig:activity-hour}b).

\begin{figure}[!tb]\centering
\includegraphics[angle=-90,width=0.48\columnwidth]{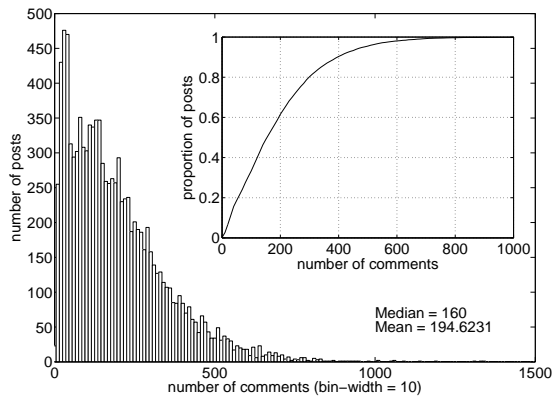}
\caption{Histogram of the number of comments per post (inset shows the corresponding cdf).}
\label{fig:hist-post}
\end{figure}
\subsection{Post-induced activity}\noindent
In this section we analyze the activity (comments) a post induces on
the site.  The histogram of Figure \ref{fig:hist-post} gives an idea
of the number of comments the posts receive.  Note that half of the
posts provoke more than $160$ comments and some of them even trigger
more than $1000$.  To analyze the time-distribution of these comments
we study their post-comment intervals (PCIs).

\begin{figure}[!t]\centering
\includegraphics[angle=-90,width=\textwidth]{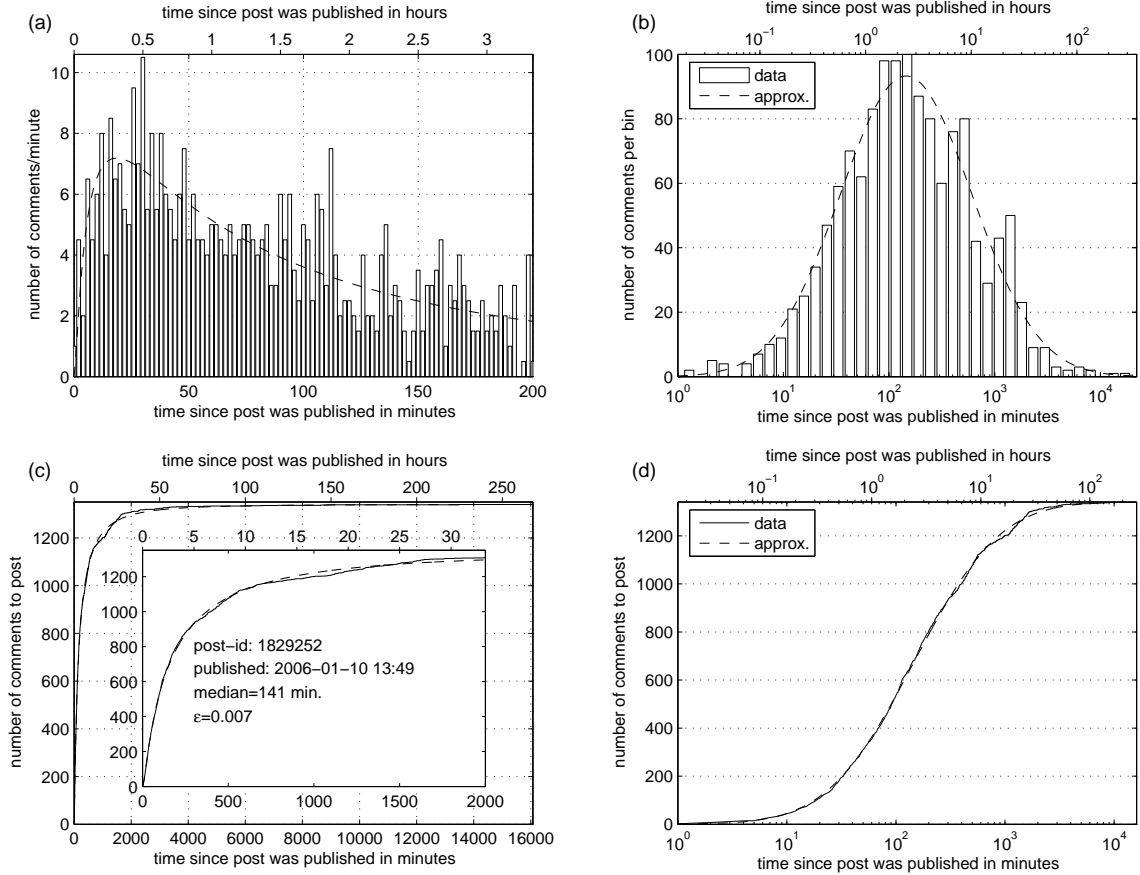}
\caption{LN-approximation (dashed lines) of the
  PCI-distribution (solid lines and bars) of a post 
   which received $1341$ comments.  
  \textbf{(a)} Comments per
  minutes (bin-with$=2$ for better visualization) for the
  first $200$ minutes after the post has been published.  
  \textbf{(b)} Same as (a) in logarithmic scale.  
  \textbf{(c)} The cumulative distribution of the data shown in (a).
  Inset shows a zoom on the first $2000$ minutes.  
  \textbf{(d)} Same as (c) in logarithmic scale.}
\label{fig:post1}
\end{figure}
\subsubsection{Analysis of the activity generated by a single post} \noindent 
We are especially interested in the resulting probability distribution
of all the PCIs of a certain post. This distribution reveals us the
probability for a post to receive a comment $t$ minutes after it has
been published.  Figures~\ref{fig:post1}a and \ref{fig:post1}b show
this distribution for a post which provoked $1341$ comments.  Although
there are some important fluctuations, the characteristic shape of the
probability density function (pdf) resembles a LN-distribution.  This
becomes even clearer if the cumulative probability distribution (cdf)
is observed, since there the fluctuations of the pdf are averaged out.
Figures~\ref{fig:post1}c and \ref{fig:post1}d show a good fit of the
PCI-cdf of the data with the cdf of the LN-distribution.  To quantify
the quality of the fit we have used a normalized error measure
$\epsilon$ based on the $\ell^1$-norm (see Appendix \ref{sec:error}).
For the post shown in Figure~\ref{fig:post1} we obtain
$\epsilon=0.007$, meaning that the average error is below $1\%$.

\begin{figure}[!tb]\centering
  \includegraphics[angle=-90,width=0.48\columnwidth]{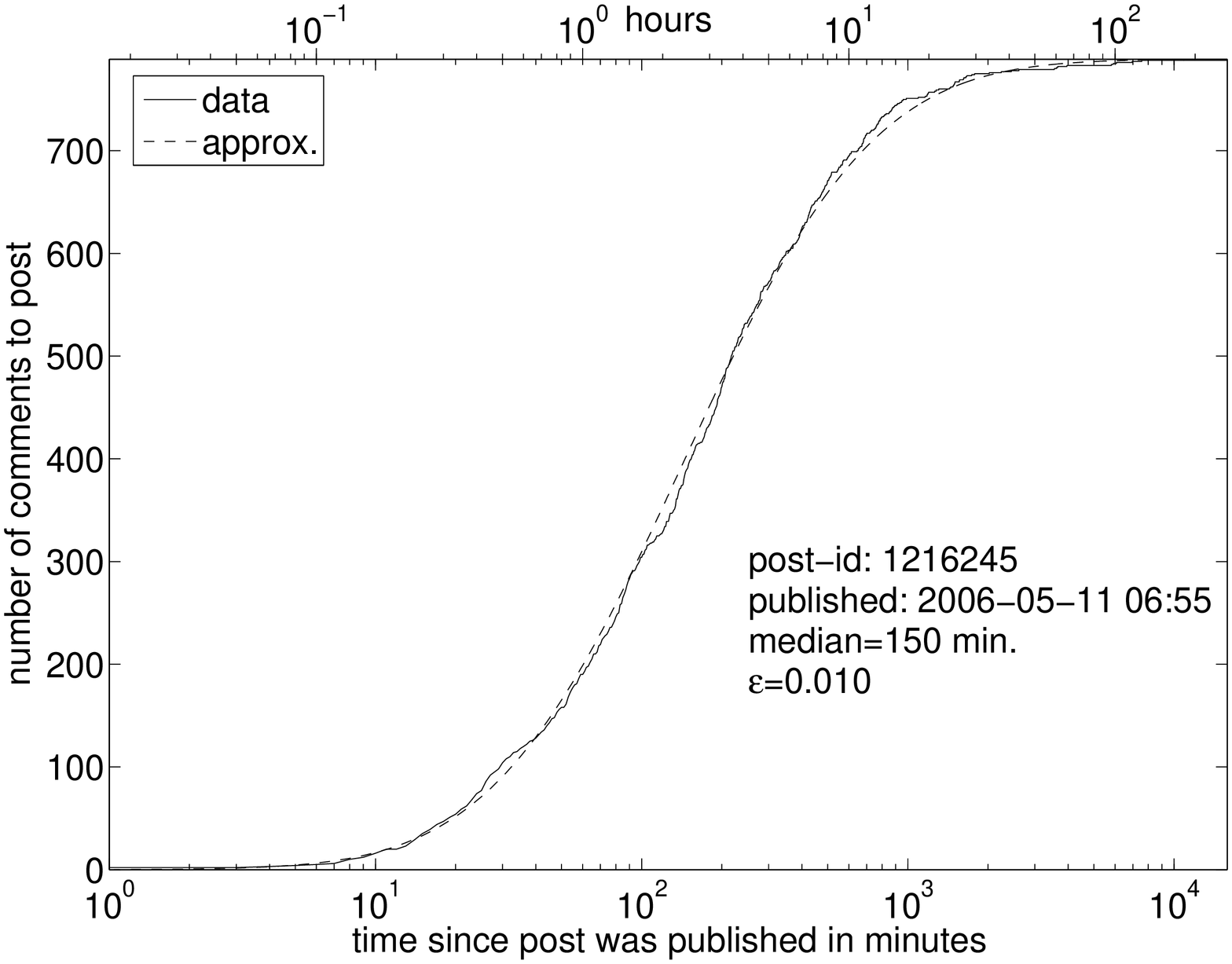}
  \includegraphics[angle=-90,width=0.48\columnwidth]{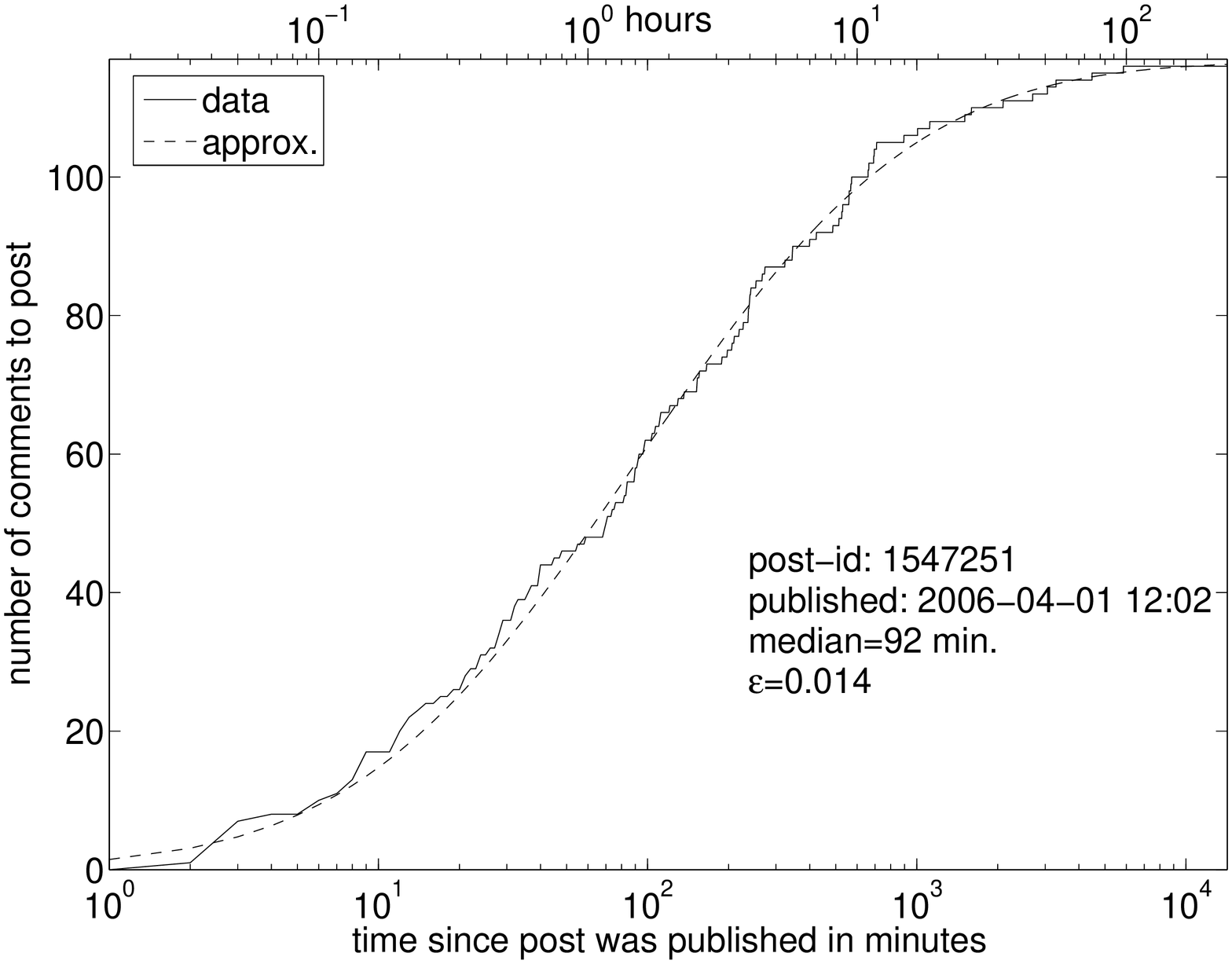}
  \includegraphics[angle=-90,width=0.61\columnwidth]{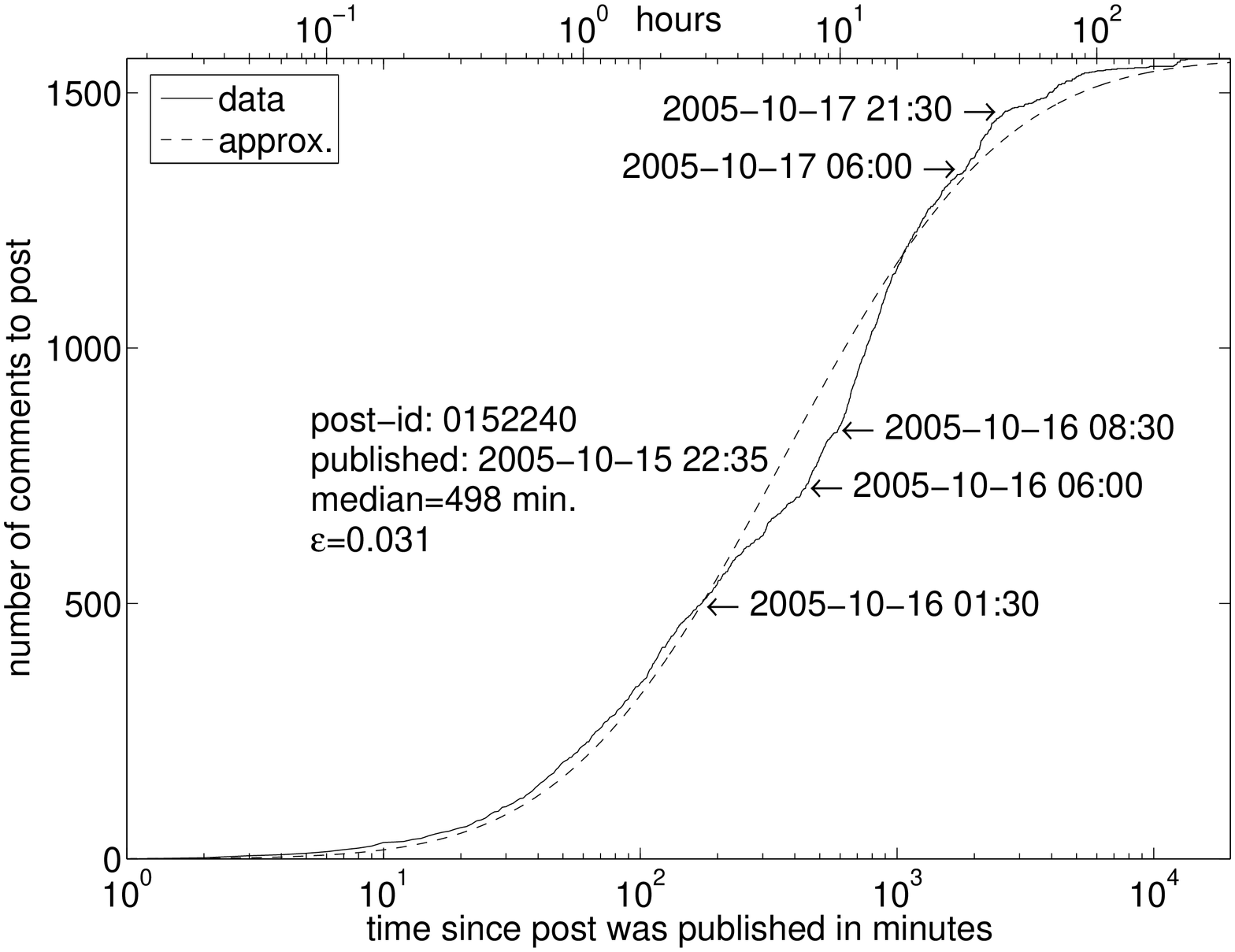}
  \caption{LN-approximation of the PCI-distribution of $3$
    different posts.}
\label{fig:posts}
\end{figure}

The PCI-cdf of three more posts can be observed in
Figure~\ref{fig:posts}. The top two sub-figures show good fits,
indicating that the PCI is well approximated even for a small number
of comments.  However, the fit is not that accurate for all posts.
E.g. the comments of the post shown in Figure~\ref{fig:posts} (bottom)
start to show considerable different behavior from the expected
LN-approximation about $3$ hours after its publication.  The activity
is lower than predicted, but starts to increase again at about $6$am
in the morning the next day. At around $8$:$30$pm it increases further
to recover the lost activity during the night. More such oscillations
of activity can be observed during the following days. The time-spans
of variations in activity coincide quite exactly with the average
daily activity cycle shown in Figure~\ref{fig:activity-hour}b.  We
analyze this coincidence further in the next section.
 \begin{figure}[!t]\centering
   \includegraphics[angle=-90,width=0.48\columnwidth]{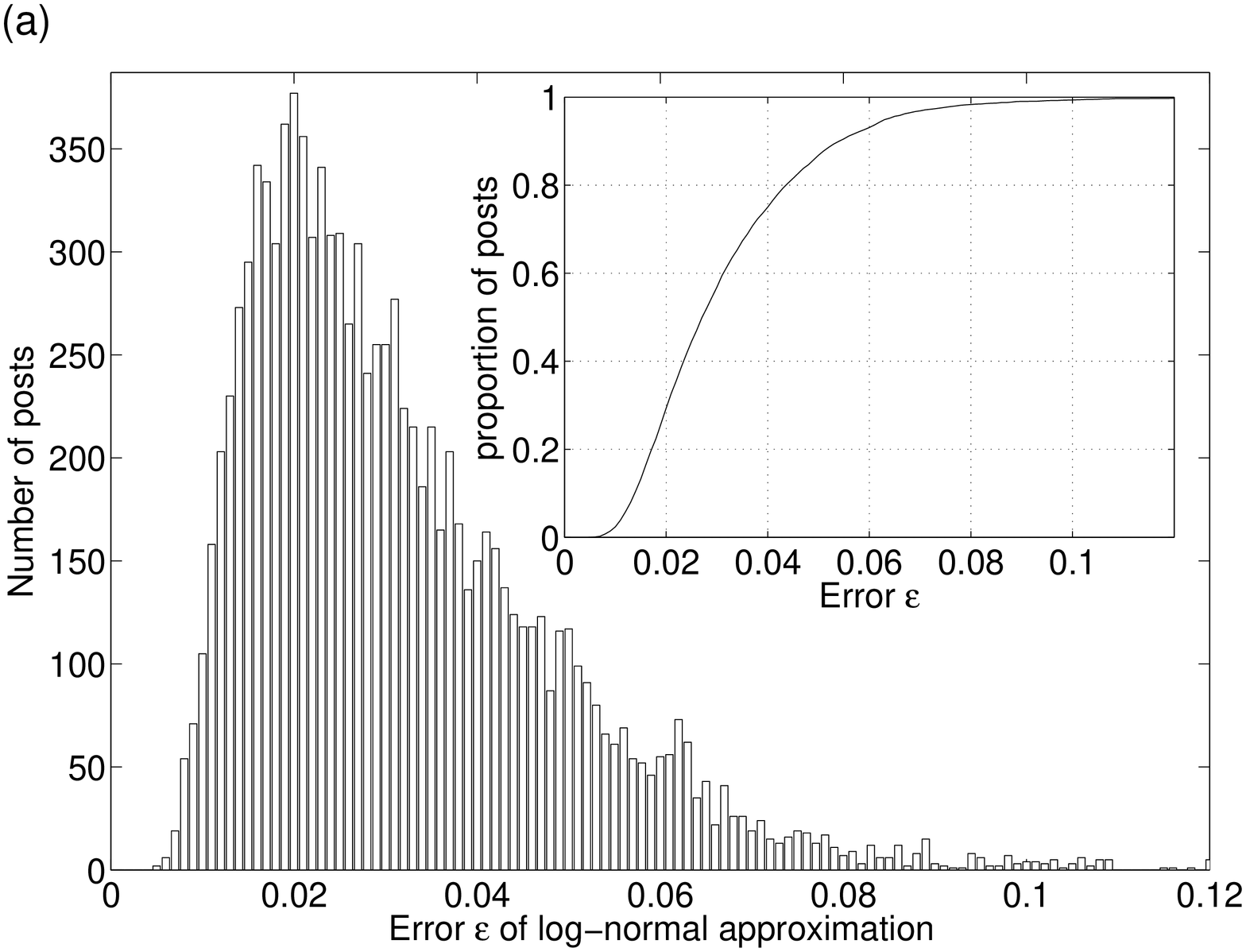}
   \includegraphics[angle=-90,width=0.48\columnwidth]{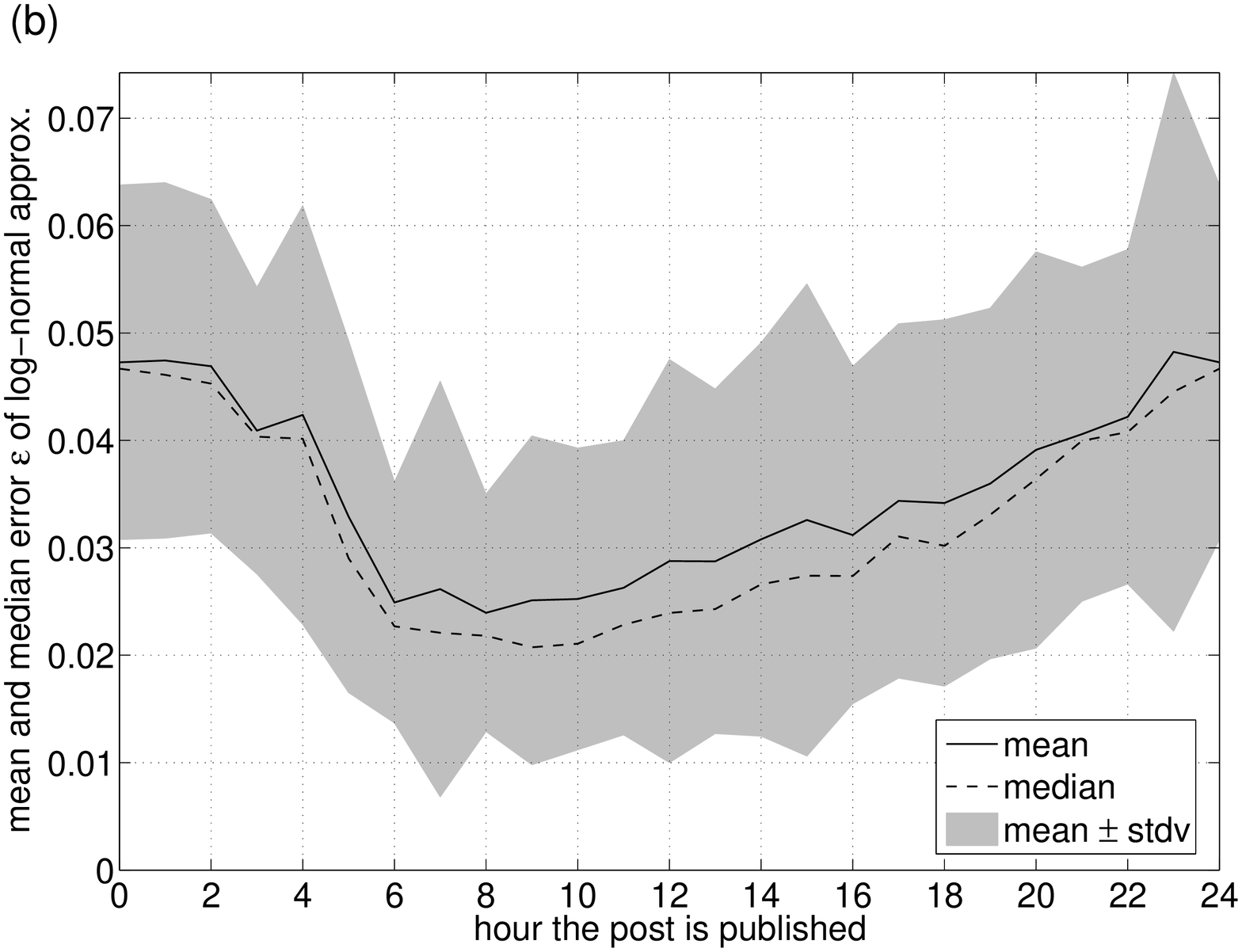}
   \caption{\textbf{(a)} Errors $\epsilon$ of the LN-approximation of
     the PCI-cdf (bin-width $=10^{-3}$). Inset shows the corresponding
     cdf.  
     \textbf{(b)} Dependence of mean and median of the approximation
     error $\epsilon$ on the hour the post is published.}
\label{fig:hour_failure}
\label{fig:hist_failure}
\end{figure}
\subsubsection{Approximation quality}\noindent
With the LN shape of the PCI-distribution identified, we focus on the
quality of this approximation in general.  We therefore calculate the
error measure $\epsilon$ of the fit for all posts which received
comments.  The resulting distribution of $\epsilon$ can be seen in
Figure~\ref{fig:hist_failure}a.  For $87$\% of the posts the
approximation error $\epsilon$ is lower than $0.05$, and for $29$\% of
them lower than $0.02$.

If we take a closer look at the data, we notice a dependence of
$\epsilon$ on the publishing-hour of a post (Figure
\ref{fig:hour_failure}b).  The best fit is reached when the post is
published between $6$am and $11$am. Then the mean error increases
successively until $11$pm to stay high during the night and recover
again in the early morning. This behavior can be understood looking at
the daily activity cycle (Figure \ref{fig:activity-hour}b).  The less
time the community has to comment on a post during the time-window of
high activity, the greater is the need to comment on it the next time
the high activity phase is reached, and hence the expected LN behavior
is altered. Figure \ref{fig:posts} (bottom) gives an example of such a
late post (published at $10$:$35$pm).

\subsubsection{Approximation with  double log-normal distributions}
\noindent 
We approximate the data as well with a double log-normal distribution
(DLN), i.e. a superposition of two LN-distributions (See appendix
\ref{sec:densities}).  To find their parameters and especially their
mixing coefficient, we use maximum likelihood estimation
\citep{Stouffer06, DeGroot2002}.  The DLN should lead to better results
in general and reduce the dependency on the circadian rhythm since it
represents two waves of activity: one starting when the post is
published and another being caused by the next increase of activity in
the circadian cycle.

\begin{figure}[!t]\centering
\includegraphics[angle=-90,width=\textwidth]{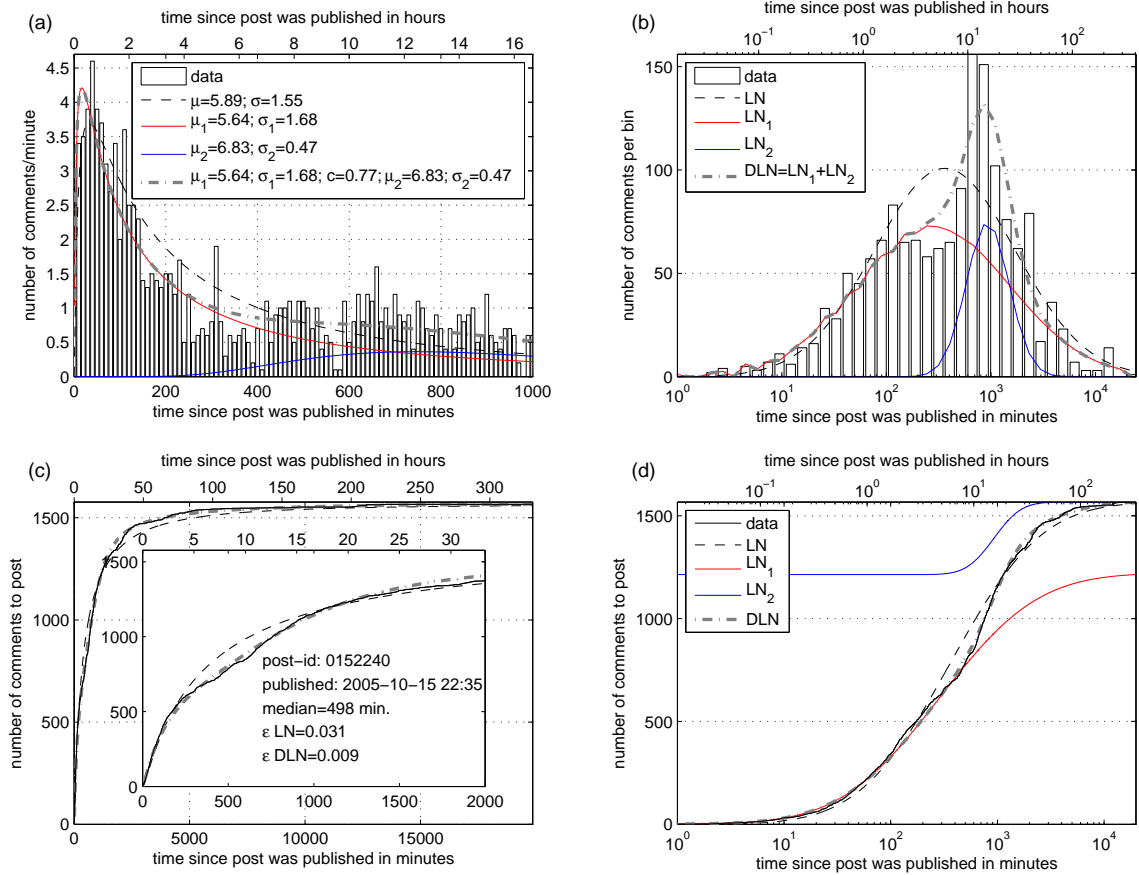}
\caption{Comparison of LN and DLN-approximations (dashed-dotted lines)
  of the PCI-distribution (solid lines and bars) of a post which
  received $1567$ comments.  The DLN-distribution is a superposition
  of LN$_1$ and LN$_2$, which in the above figure are rescaled
  according to the coefficient $c$ of the DLN. Rest of legend as in
  Figure \ref{fig:post1}}.
\label{fig:post_DLN}
\end{figure}

An example of this behavior is shown in Figure \ref{fig:post_DLN}
where we compare LN and DLN-ap\-prox\-i\-ma\-·tion of the same post as used in
Figure \ref{fig:posts} (bottom).  The red and blue lines indicate the
two log-normals whose superposition results in a DLN (gray,
dashed-dotted), which clearly outperforms the previous LN (black,
dashed) approach.  The error $\epsilon$ decreases from $0.031$ to
$0.009$ and the approximation is much closer to the cdf of the data
(black continuous line in Figures \ref{fig:post_DLN}c and
\ref{fig:post_DLN}d). We notice that the first $10$ hours of activity
are well approximated by a single LN-distribution (red line).
Then the activity increases due to the high phase of the circadian
cycle (compare also with the labels of Figure \ref{fig:posts} bottom).
The second LN distribution (blue line) accounts for this
increase and therefore the DLN-approximation reflects the first bump
in the PCI-cdf and fits well the data.

\begin{figure}[!tb]\centering
  \includegraphics[angle=-90,width=0.48\columnwidth]{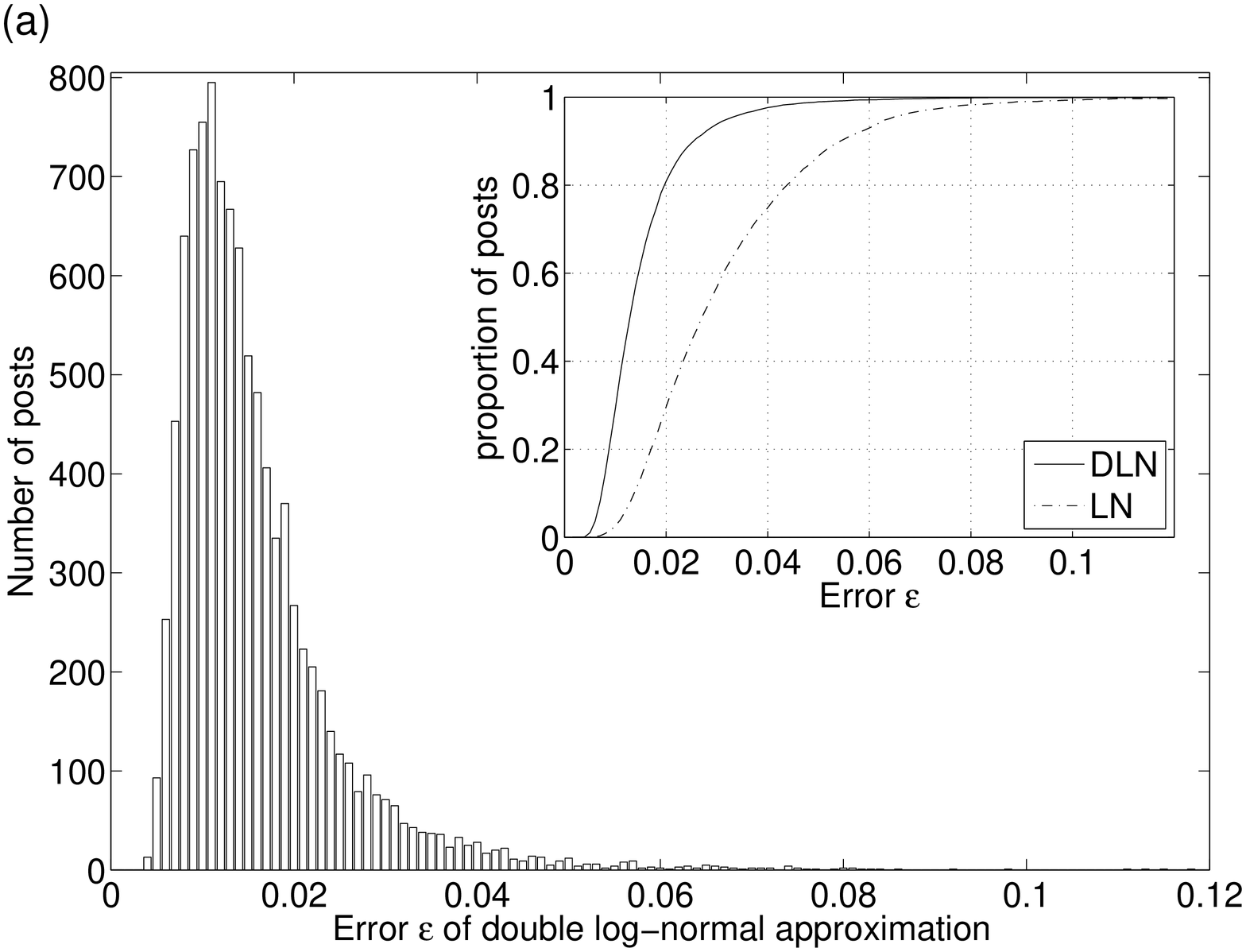}
  \includegraphics[angle=-90,width=0.48\columnwidth]{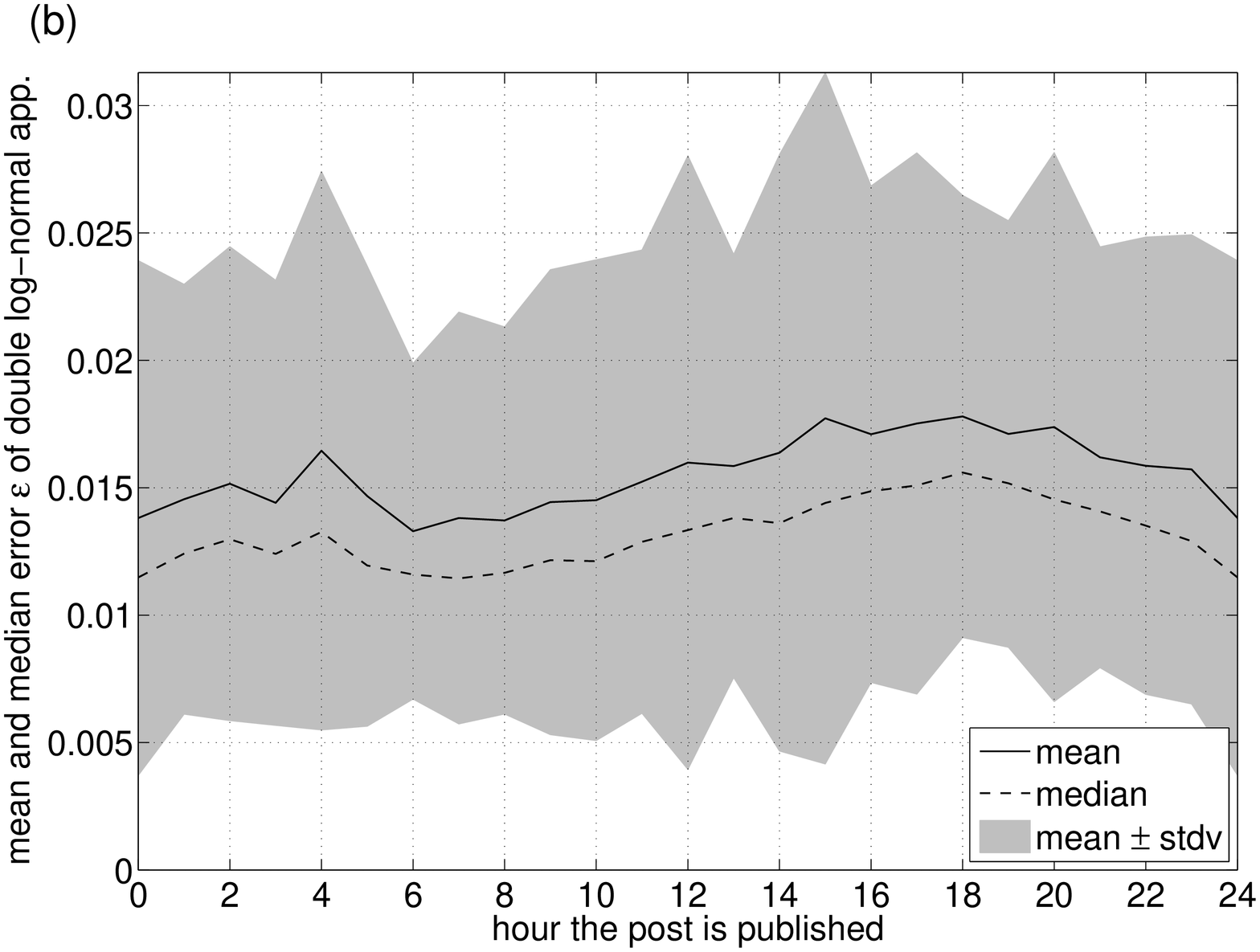}
  \caption{\textbf{(a)} Errors $\epsilon$ of the DLN-approximation of
    the PCI-cdf (bin-width $=10^{-3}$). Inset shows the corresponding
    cdf.  \textbf{(b)} Dependence of mean and median of the
    approximation error $\epsilon$ on the hour the post is published.}
\label{fig:hour_failureDLN}
\label{fig:hist_failureDLN}
\end{figure}

To quantify the overall performance of a DLN-fit we apply it on all
posts and plot the distribution of its approximation error $\epsilon$
in Figure \ref{fig:hist_failureDLN}a. The inset compares the
error-cdfs of DLN (continuous) and LN-approach (dashed-dotted). We
notice a significant improvement of the approximation quality. For
example, the error of the DLN-fits is below $0.02$ for more than
$80\%$ of the posts compared to only $29\%$ in the case of
LN-approximations.  Figure \ref{fig:hour_failureDLN}b shows only a
minor dependency of the quality of the DLN-fits on the publishing hour
of the post (compare with Figure \ref{fig:hour_failure}b), which
allows us to conclude that the DLN-distributions accounts for the
major part of the aberration of the log-normal behavior caused by the
circadian cycle.

\subsubsection{Approximation parameters}\noindent
For the cases where a LN-distribution leads to good results we can
describe the activity triggered by a post with only two parameters:
the median\footnote{Note that the median coincides with the geometric
  mean for a log-normally distributed random variable.}  and the
geometric standard deviation $\sigma_g$ of the PCI-pdf, commonly used
to compare log-normally distributed quantities~\citep{LimpertSA01}.
The median and $\sigma_g$ relate to the parameters of the
LN-distribution in the following way.
\begin{align}
\mbox{median}&=\exp(\mu)~,  &\sigma_g&=\exp(\sigma).
\label{eq:geo}
\end{align}

\begin{figure}[!htb]\centering
  \includegraphics[angle=-90,width=0.48\columnwidth]{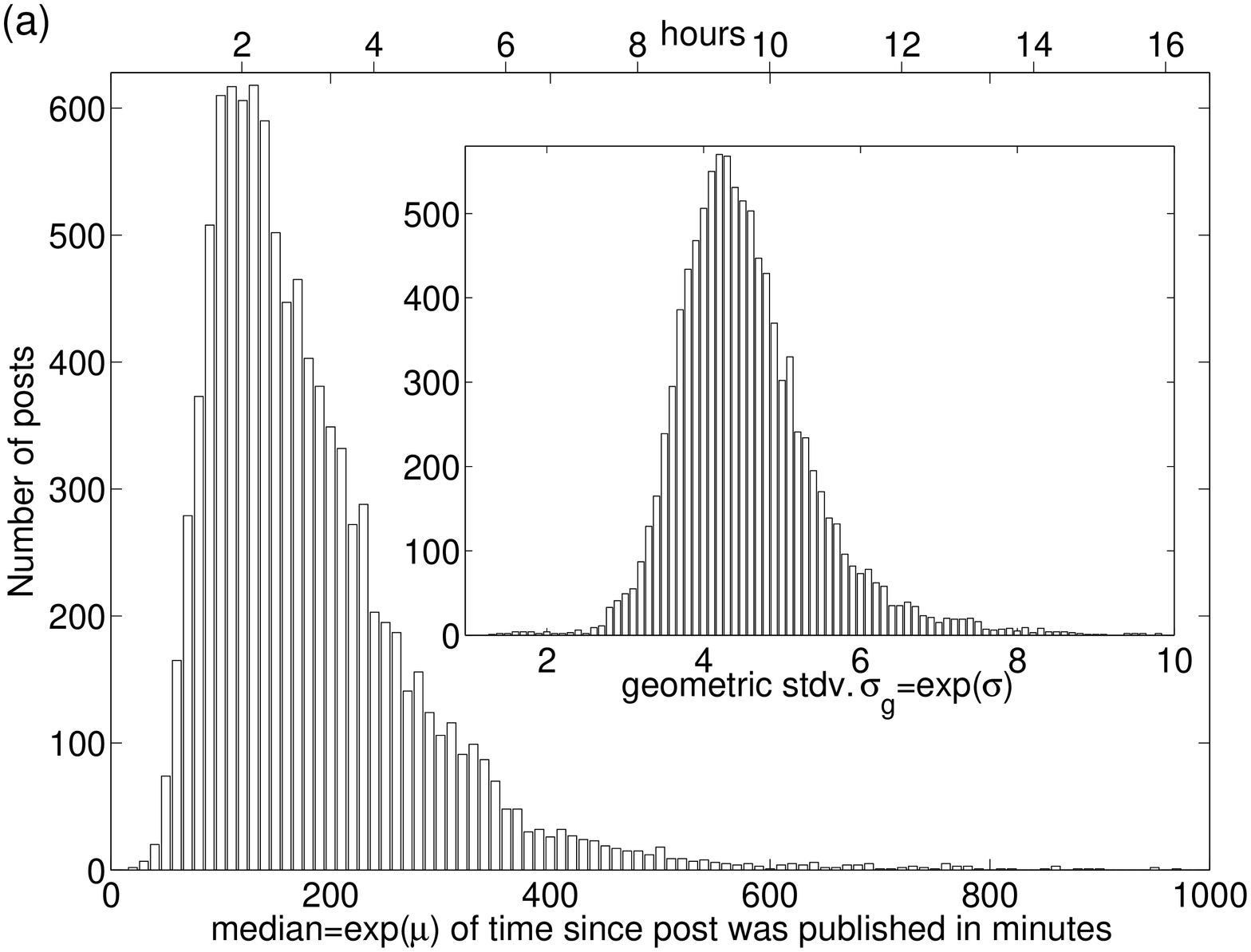}
  \includegraphics[angle=-90,width=0.48\columnwidth]{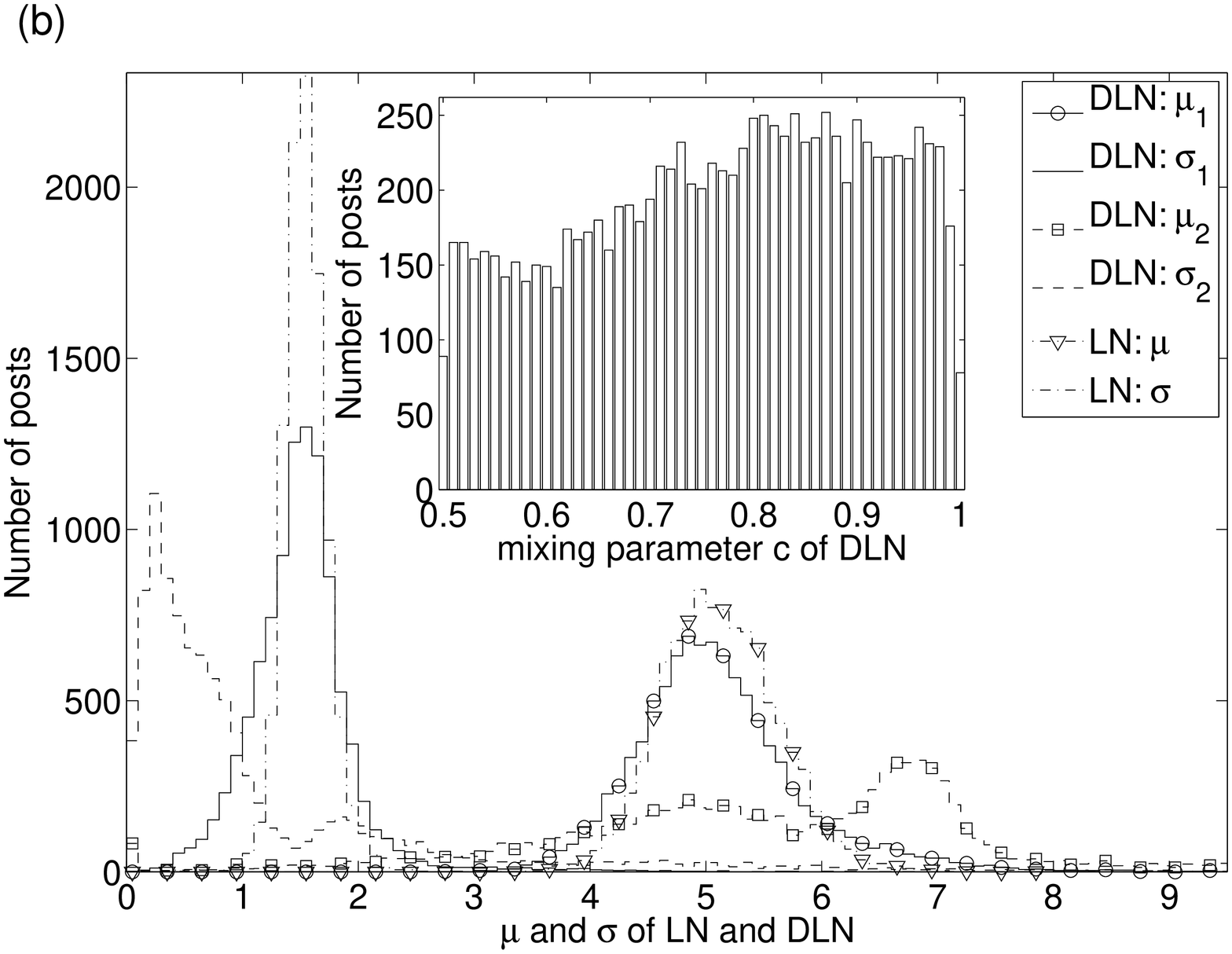}
  \caption{\textbf{(a)} Histograms of the estimates of medians
    (bin-width $=10$) and geometric standard deviations (inset,
    bin-width $=0.1$) of the PCI-distributions.  \textbf{(b)}
    Parameters of LN and DLN-approximations.  Bin-width=$0.1$ for
    $\mu$ and $\sigma$, $0.01$ for $c$ (inset).  }
\label{fig:hist_median}
\end{figure}

Figure~\ref{fig:hist_median}a shows the distribution of these
quantities for all posts\footnote{Instead of calculating $\sigma_g$
  directly from the data as in a previous version of this study
  \citep{kaltenbrunner_saw2007} we used equation \eqref{eq:geo} and the
  estimates of $\sigma$, which led to different results. Compare also
  with~\citet{LimpertSA01}.}.  The inset shows the distribution of
$\sigma_g$, which is centered around $4.5$ and has a standard
deviation of $0.91$. The median of the post-induced activity on the
other hand shows more variations, but is rather short (for $50$\% of
the posts it is below $2.5$ hours, for $90$\% below $6$ hours)
compared to the maximum PCI (approx. $12$ days). We can thus conclude
that although the total activity a post generates covers a large time
interval, the major part of the activity happens within the first few
hours after the post's publication.

If we use a DLN-distribution to approximate the data we need five
parameters. Their distributions together with those of the parameters
$\sigma$ and $\mu$ of the LN-approximation are displayed in Figure
\ref{fig:hist_median}b. For better visualization we choose a stair
plot instead of a bar-graph. Clearly the regions of $\mu_1$
(continuous line with circles) and $\sigma_1$ (continuous line) are
very similar to those of the parameters of LN-approximations
(dashed-dotted lines), indicating that the first one of the two
log-normal distributions used to generate the DLN is similar to the
LN-approximations.  The parameters $\mu_2$ and $\sigma_2$, on the
other hand, show an interesting bimodal behavior.  One of the two
peaks of the distribution falls within the regions of $\mu_1$ or
$\sigma_1$ respectively. Those cases correspond to posts for which the
two superposed log-normal distributions are very similar and the data
fits well already a single LN-distribution.  The second peak in the
$\mu_2$-distribution represents those posts which provoke a second
wave of activity due to the circadian cycle.  In those cases the
parameter $\sigma_2$ is usually smaller than $\sigma_1$.  The inset of
Figure \ref{fig:hist_median}b shows the mixing parameter $c$, which is
nearly uniformly distributed although values in $[0.7,1]$ are slightly
more likely than lower ones.  We sorted the parameters to ensure a
value of $c\geq0.5$.

\subsection{User dynamics}\noindent
In this section we analyze the activity on Slashdot taking the
authorship of the comments into account. We first study the
distribution of activity among all the users participating in the
debates and then focus on the temporal activity patterns of single
users.

\subsubsection{Global user activity} 
\begin{figure}[!t]\centering
  \includegraphics[angle=-90,width=0.48\columnwidth]{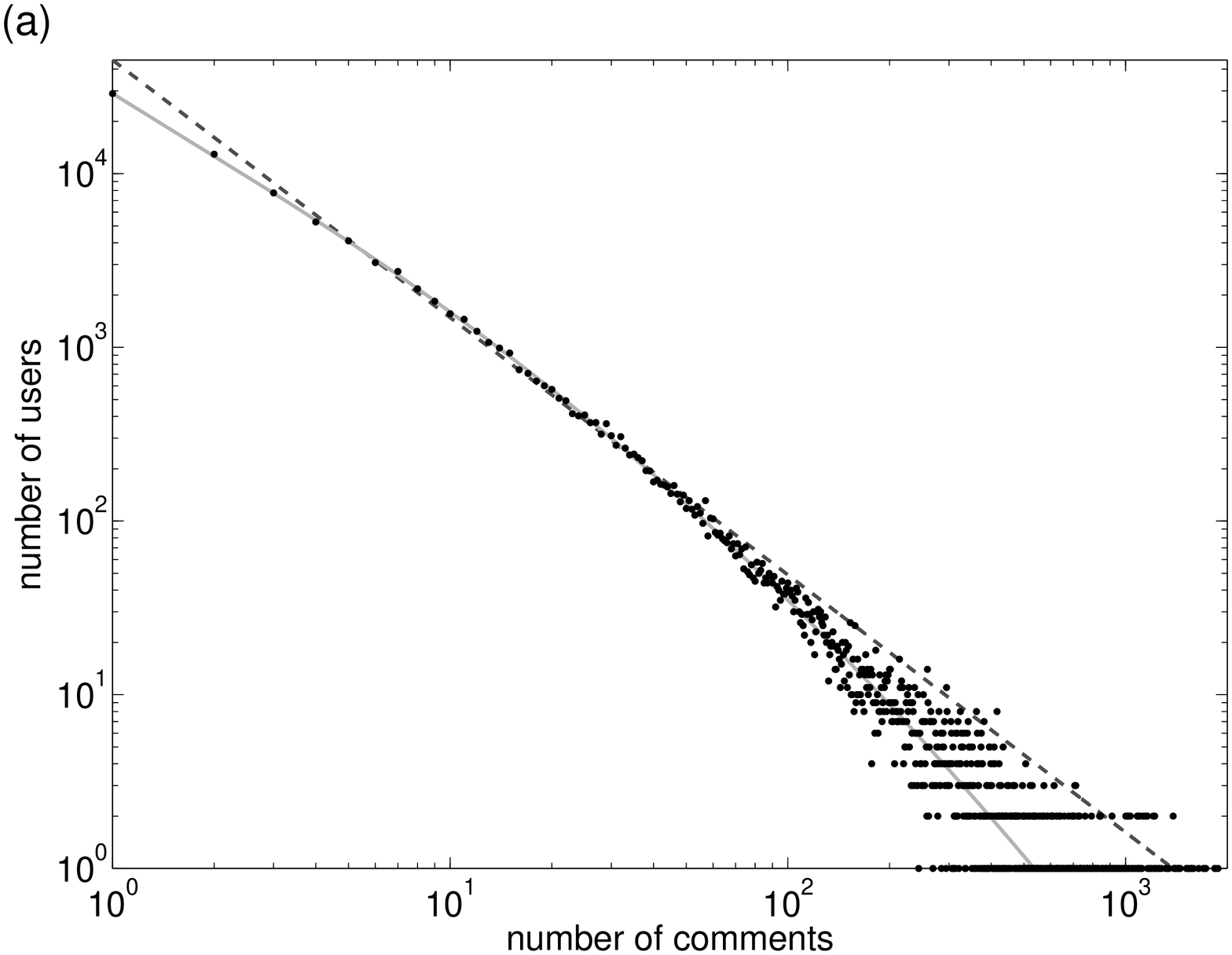}
  \includegraphics[angle=-90,width=0.48\columnwidth]{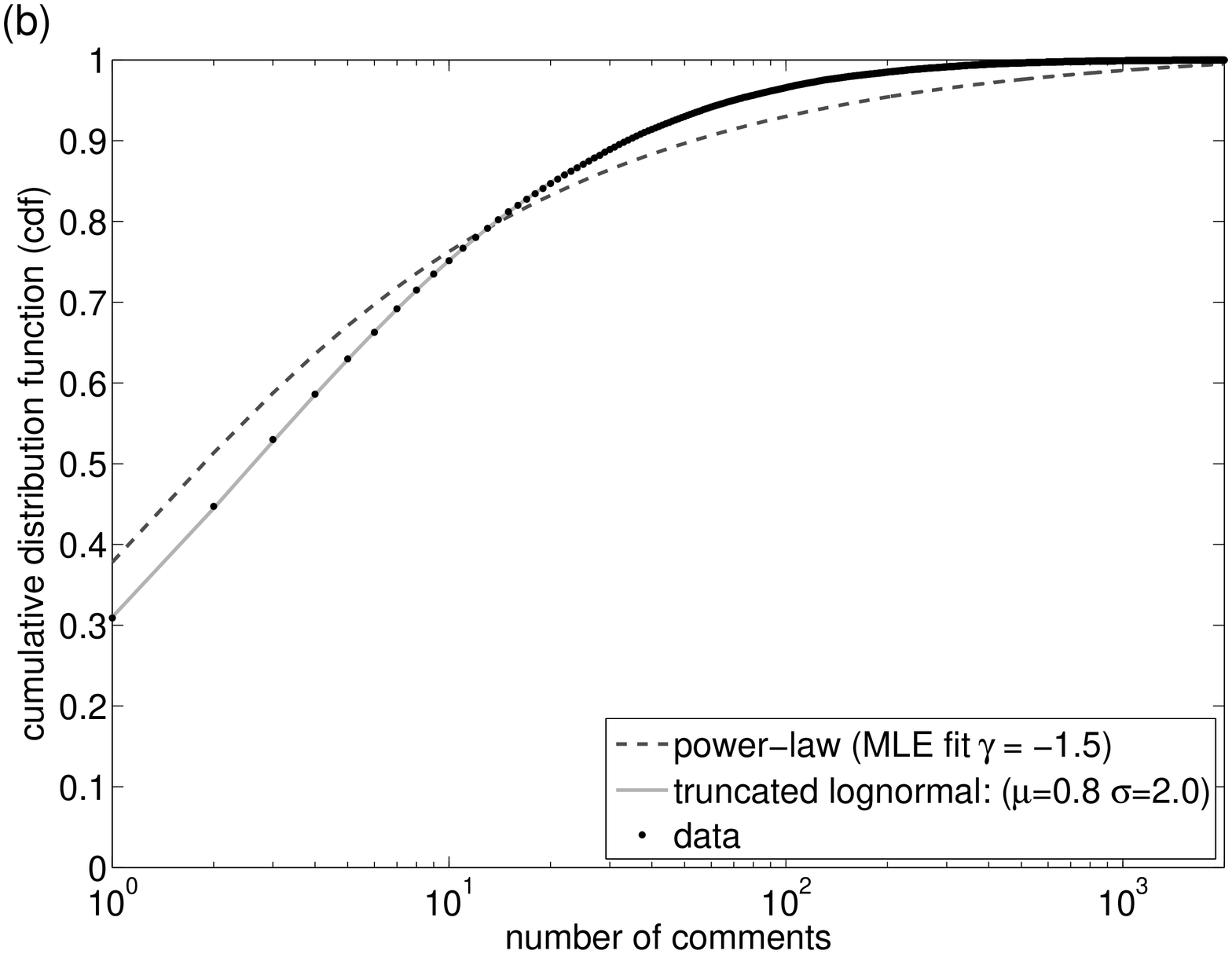}
  \caption{\textbf{(a)} Histogram of the number of comments per user
    and \textbf{(b)} and its corresponding cdf.}
\label{fig:powerlaw}
\end{figure}

\noindent 
The activity of all users is best illustrated by the distribution of
the number of comments per user. It is shown in double-logarithmic
scale in Figure~\ref{fig:powerlaw}a.  The obtained distribution
follows quite closely a straight line, suggesting a power-law
probability distribution governing this relation.  We note that $53\%$
of the users write $3$ or less comments whereas only $93$ users
($0.1\%$) write more than $1000$ comments.  Indeed, after applying
linear regression as in other studies
\citep{faloutsos1999, barabasi1999} we obtain a quite large correlation
coefficient $R^{2}=-0.97$ for an exponent of $\gamma = -1.79$.

However, if we apply rigorous statistical analysis as proposed by 
\citet{goldstein2004} the picture changes.  First,
we estimate the power-law exponent computing the less biased maximum
likelihood estimator (MLE).  The resulting exponent $\gamma = -1.5$
differs significantly from the previous one and is illustrated in
Figure~\ref{fig:powerlaw} (dashed-line).  Although
Figure~\ref{fig:powerlaw}a tempts one to accept the power-law
hypothesis, the cdf shown in Figure~\ref{fig:powerlaw}b discards it.
It is thus not surprising that the Kolmogorov-Smirnov test forces us
to reject the power-law hypothesis with statistical significance at
the $0.1\%$ level.

As an alternative hypothesis to describe the data we propose a
truncated LN probability distribution, shown in
Figure~\ref{fig:powerlaw} as grey-solid-line. Its parameters are found
using the MLE.  Clearly, the fit is better using this hypothesis.  We
remark that in many studies some data points (considered outliers) are
discarded to improve the power-law fit.  Here, in contrast, the
truncated LN-approximation can characterize the entire data-set.

\subsubsection{Single user dynamics} \noindent 
After characterizing the user activity at a general level, we
investigate the temporal behavior patterns of single users . The
analysis concentrates on the two most active users (to protect their
privacy we call them user$1$ and user$2$). Table \ref{table:active}
shows the number of commented posts and the total number of comments
these two users published during the time-span covered by our data.
\vspace{-12pt}
\begin{table}[!hb]\centering
\caption{Contributions of the two most active users.}
\begin{tabular}{lrr} 
& user1                     & user2\\
\hline 
commented posts & $1189$ & $1306$ \\
comments        & $3642$ & $3350$ 
\end{tabular}
\label{table:active}
\end{table}

\noindent
We focus on the distribution of the PCIs of all of their comments as
well as on their inter-comment-interval (ICI) distribution, i.e.  the
time-difference between two comments of the same user.

\begin{figure}[!t]\centering
  \includegraphics[angle=-90,width=0.48\columnwidth]{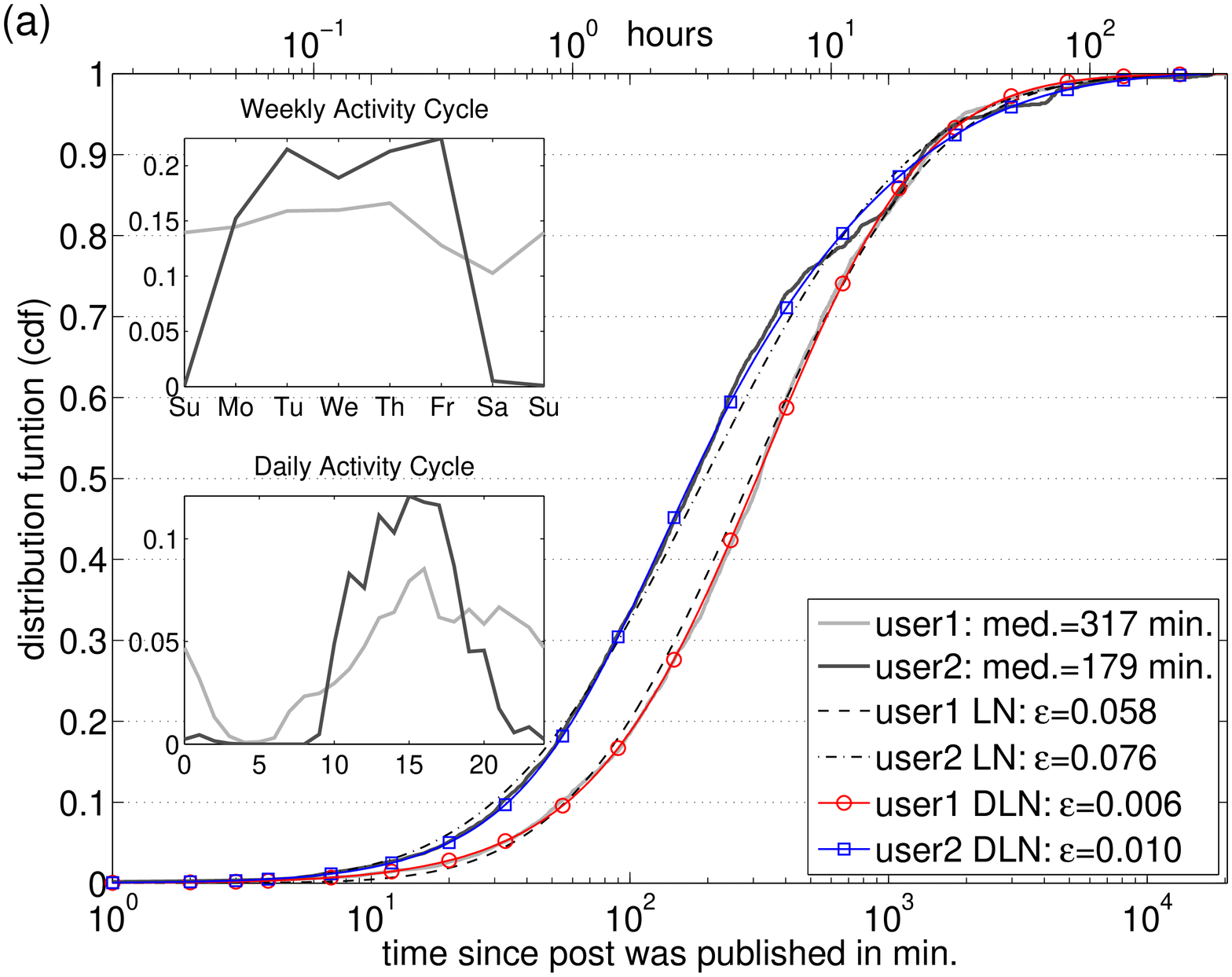}
  \includegraphics[angle=-90,width=0.48\columnwidth]{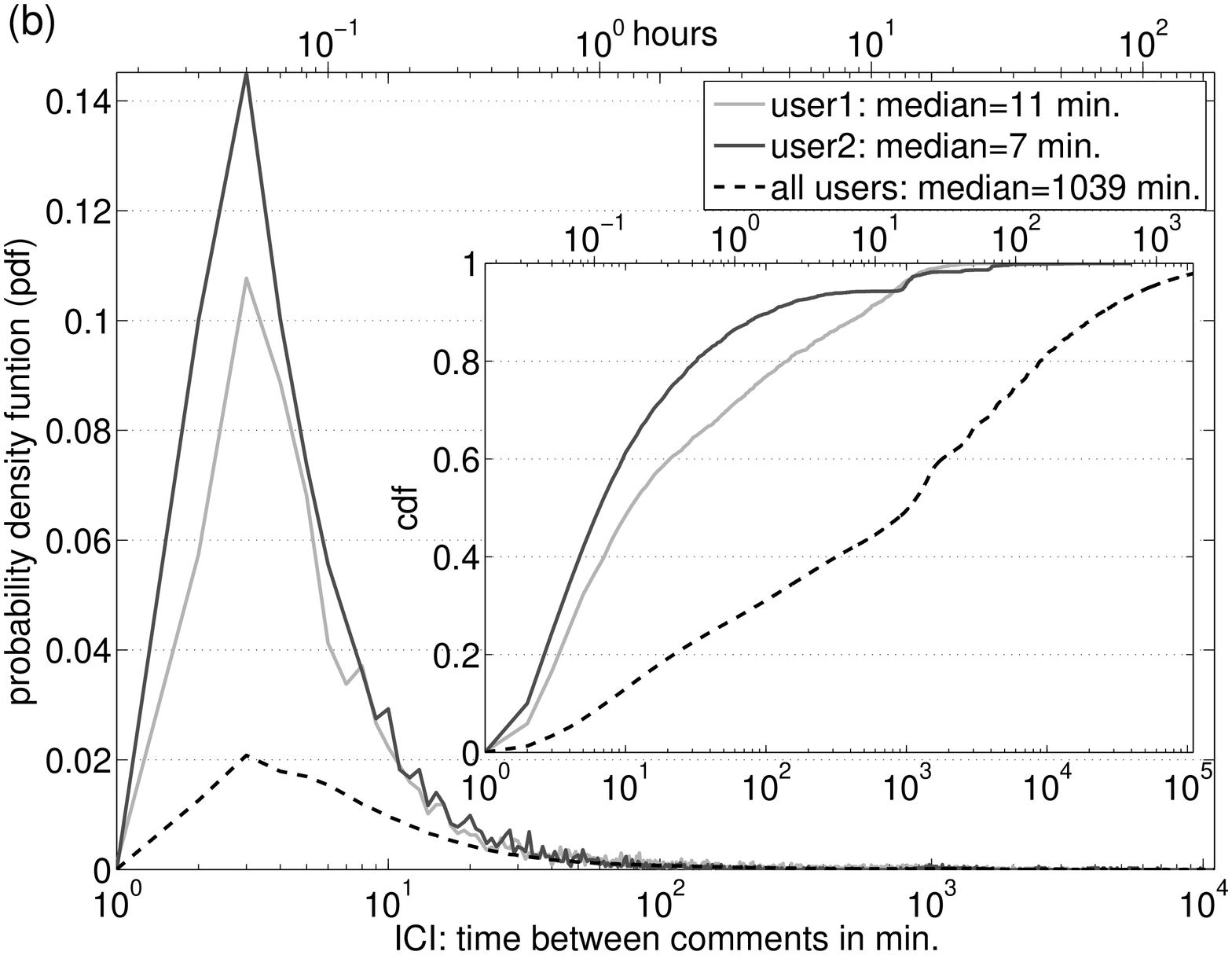}
  \caption{Activity patterns of the two most active users:
    \textbf{(a)} PCI-distributions, insets shows daily and weekly
    activity cycles.  
    \textbf{(b)} Distribution of the inter-comment intervals (ICI)
    compared with the whole population (dashed line).}
\label{fig:PCI_active}
\label{fig:ICI_active}
\end{figure}

We approximate the PCI-cdf (gray lines in Figure
\ref{fig:PCI_active}a) also with LN (dashed and dashed-dotted lines)
and DLN-distributions (blue and red lines with box and circle
markers). The quality of the LN-fit is worse than in the case of the
post-induced comment activity, but the DLN-distribution is a good
explanation of the data with a small approximation error $\epsilon$.
Again we notice a clear dependence of the quality of the fit on the
activity cycle (shown in the insets of Figure \ref{fig:PCI_active}a).
The approximation is much better for user1, whose daily and especially
weekly activity cycles are much more balanced than those of user2. The
activity of the latter user concentrates almost exclusively on the
working hours from Monday to Friday.  Hence his PCI-distribution shows
a clear decrease after $8$ but increases again after $16$ hours. This
increase is less pronounced if only the first comment to a post is
considered (data not shown), indicating that the user frequently
rechecks the posts he commented the day before to participate again in
an ongoing discussion.

The same effect can be observed in their ICIs, which are illustrated
in Figure \ref{fig:ICI_active}b.  There the cdf (inset of Figure
\ref{fig:ICI_active}b) of user$2$ shows an even more pronounced
increase around an ICI of $16$ hours. We further observe that the
ICI-pdf peaks for both users as well as for the whole population at
$3$ minutes. This is probably caused by an anti-troll
filter~\citep{SlashdotFAQ}, which should prevent a user from commenting
more than once within $120$ seconds.  The medians of the
ICI-distributions of user1 and user2 are rather short ($11$ and $7$
minutes respectively) compared to the median of the whole population
(about $17$ hours), indicating that the two users engage in
discussions frequently during their activity phase.

\section{Discussion}
\label{sec:discussion}\noindent
The special architecture of the technology-related news website
Slashdot allowed us to analyze the temporal communication patterns of
an online society without considering semantic aspects. The site
activity is driven by news-posts which provoke communication activity
in the form of comments.

Despite the great amount of users participating in the discussions,
close to $10^5$ in the data we have studied, and the diversity of
themes (games, politics, science, books, etc.) some simple patterns
can be identified, which repeat themselves over and over again. One of
these patterns appears in the shape of the distribution of time
differences between a post and its comments (the PCIs). It can be well
approximated by a log-normal distribution (Figures~\ref{fig:post1} and
\ref{fig:posts}) for most of the posts. The only remarkable deviations
from these approximations are caused by oscillatory daily and weekly
activity patterns (Figure~\ref{fig:activity-hour}), which become less
noticeable if a post is published early in the morning
(Figure~\ref{fig:hour_failure}a). A significant improvement of the
approximation can be achieved using a superposition of two log-normal
distributions. Such a double log-normal accounts for the first
oscillation caused by the circadian cycle.  It can be interpreted as
two independent waves of activity, one starting directly after a post
has been published, and the second at the next increase of activity
due to the circadian rhythm. Although more such oscillations may occur
during the life-time of a post, their amplitude is low compared to the
first one, suggesting that a combination of more than two
LN-distributions would only increase the complexity of
parameter-finding (via MLE) without improving significantly the
approximation quality.  Nevertheless, a combination of a
DLN-distribution with an oscillatory function emulating the circadian
cycle leads to slightly better results \citep{kaltenbrunner_LAWEB2007},
without affecting the complexity of MLE.

In single user behavior an akin pattern appears in the
PCI-distribution of all of the comments a user writes to several posts
(Figure \ref{fig:PCI_active}a). Again deviations are caused by the
circadian cycle.  Another interesting pattern can be observed
analyzing the ICI of single-users, i.e. the time-span between two
consecutive comments of a certain user.  In the case of the two most
active users (Figure~\ref{fig:PCI_active}b) the ICI-distributions are
very similar, which further supports our hypothesis of the existence
of homogeneous temporal patterns on Slashdot.

We would expect that the time-spans between publishing and reading of
a post also follow log-normal patterns. This could be easily verified
checking the server logs of Slashdot or access-times of an external
homepage linked by a Slashdot post.  Such a study has been performed
to show the Slashdot effect~\citep{Adler1999}, but the scale of the
data presented does not allow to draw significant conclusions. Further
investigation is needed to verify this claim.

Log-normal temporal patterns similar to those described above were
found in person-to-person communication by \citet{Stouffer06},
who investigated the waiting and inter-event
times of an e-mail activity dataset. A second coincidence between
their study and our findings is that the number of comments (or
e-mails in their case) can be well approximated by the same
distribution (a truncated log-normal in this case).  The temporal
patterns of the e-mail data were previously claimed to show power-law
behavior, which would be explained by a queuing model
\citep{barabasi05}.  Although this model might allow insight into other
types of human activity~\citep{Vazquez06} it is not able to account for
the observed log-normal behavior patterns. We hope therefore to
encourage further research towards a theoretical understanding of the
underlying phenomena responsible for this apparently quite general
human behavior pattern.

The medians (Figure~\ref{fig:hist_median}) of the PCI-distributions
are very small compared to the overall duration of the activity
provoked by a post.  Although the posts might be available for
commenting for more than $10$ days, the first few hours decide whether
they will become highly debated or just receive some sporadic
comments.  We would therefore expect that the simplicity of the
approximation together with the high initial activity should make an
accurate prediction of the expected user behavior feasible at an early
phase after a post has been put online.  The accuracy of such
forecasting methods is subject of current research
\citep{kaltenbrunner_LAWEB2007}.

An early characterization of the activity triggered by a post could be
applied, for instance, on dynamic pricing or placing of online
advertisements or on the improvement of online marketing. The success
of a campaign might be predicted already after a short time-period,
thus allowing an early adaptation of the strategy of information
diffusion. In this context the viral marketing concept
\citep{Leskovec06}, which relies on personal communication might be the
most promising field.

In our opinion, the regular communication activity patterns described
in this work may be relevant in two aspects. The first, simpler one,
is related to applications where a better understanding of information
trade in the web translates easily into a better description, and even
quantification, of Internet audience.  But a second, more complex,
aspect is related to the human ``communicative'' behavior uncovered at
present time: Internet based communication capabilities. We face a
new, large scale, all-to-all public space in which a novel kind of
social behavior arises, a scenario that we do not yet fully
understand. However, we should not forget that the new activity is
being largely recorded and the data can be available for research.
The work presented in this contribution is a good example of how those
data can be collected and analyzed to give, at least, a quantitative
description of the behavior.  This is a first step towards a more
ambitious target: to develop ``ab initio'' models for the population
dynamics of message interchange, which is also the goal of our current
research.

\acks{ This work has been partially funded by C\`atedra Telef\'onica de
 Producci\'o Multim\`edia de la Universitat Pompeu Fabra.}


\begin{appendix}
\section{Log-normal and double log-normal distributions}
\label{sec:densities}\noindent
The following two probability distributions have been used in this article:\\
A \textbf{log-normal (LN)} distribution, which has the following
  probability density function (pdf):
\begin{eqnarray}
  f_{LN}(t;\mu,\sigma)&=&
  \frac{1}{t\sigma\sqrt{2\pi}}\exp\left(\frac{-(\ln(t)-\mu)^2}{2\sigma^2}\right)
\label{eq:ln_pdf}
\end{eqnarray}
and its cumulative distribution function (cdf) is given by:
\begin{eqnarray}
  F_{LN}(t;\mu,\sigma)&=&\frac{1}{2}+\frac{1}{2}\mbox{erf}\left(\frac{\ln(t)-\mu}{\sqrt{2}\sigma}\right),
\label{eq:ln_cdf}
\end{eqnarray}
where $\mbox{erf(x)}$ is the Gauss error function
being defined as
\begin{eqnarray}
  \mbox{erf}(x)=\frac{2}{\sqrt{\pi}}\int_0^x\exp(-u^2)du.
\label{eq:erf}
\end{eqnarray}
And a \textbf{double log-normal (DLN)} distribution, which is a superposition
  of two independent LN-distributions and has the following pdf:
\begin{eqnarray}
  f_{DLN}(t;\theta)&=&
  c f_{LN}(t;\mu_1,\sigma_1)+(1-c)f_{LN}(t;\mu_2,\sigma_2) \label{eq:2ln_pdf}\\
\mbox{where } \theta&=&(\mu_1,\sigma_1,c,\mu_2,\sigma_2)\nonumber.
\end{eqnarray}
The corresponding cdf can be easily derived from equations \eqref{eq:ln_cdf} and \eqref{eq:2ln_pdf}.
\section{Error Measure $\epsilon$}
\noindent \label{sec:error}\hspace{-1mm} We use the following distance
measure to calculate the error of the approximations. The distance
between approximation and data is only calculated for the time-bins
(i.e.  minutes) where a post actually receives a comment to avoid a
distortion of the error measure by the periods with low comment
activity.
\begin{definition}
  Let $\mathbb{T}$ be the set of time-bins where a post receives at
  least one comment and $T$ its cardinality.  We define then the
  approximation error $\epsilon$ of a function $f(t)$ approximating
  $g(t)$ (both defined for all $t \in \mathbb{T}$) as the normalized
  $\ell^1$-norm of $f(t)-g(t)$:
\begin{equation}
\epsilon=\sum_{t \in \mathbb{T}}\frac{|f(t)-g(t)|}{T}.
\end{equation}
\end{definition}\noindent
If $f(t)$ and $g(t)$ are cumulative probability density functions
(i.e. $0 \leq f(t) \leq 1$ and $0 \leq g(t) \leq 1$), it follows that $0
\leq \epsilon \leq 1$.  
\end{appendix}

\vskip 0.2in
\bibliographystyle{hapalike.bst}
\bibliography{mybiblio}

\end{document}